\documentclass[aip,jmp,reprint,floatfix]{revtex4-1}
\usepackage{graphicx}
\usepackage[colorlinks,allcolors=blue]{hyperref}
\usepackage{amsmath,amsfonts}
\newcommand{\pr}{\mathbb{P}}

\newcommand{\dpar}[1]{\left(#1\right)}
\newcommand{\dsqr}[1]{\left[#1\right]}
\newcommand{\dcur}[1]{\left\{#1\right\}}
\newcommand{\dabs}[1]{\left|#1\right|}
\newcommand{\floor}[1]{\lfloor#1\rfloor}

\newcommand{\beq}{\begin{equation}}
\newcommand{\eeq}{\end{equation}}

\begin{document}
\title{Convergence of the probability of large deviations in a model of correlated random variables having compact-support \texorpdfstring{$Q$}{Q}-Gaussians as limiting distributions}


\author{Max Jauregui}
\email{jauregui@cbpf.br}
\affiliation{Centro Brasileiro de Pesquisas F\'isicas and National Institute of Science and Technology for Complex Systems, Rua Xavier Sigaud 150, Rio de Janeiro 22290-180, RJ, Brazil}
\author{Constantino Tsallis}
\email{tsallis@cbpf.br}
\affiliation{Centro Brasileiro de Pesquisas F\'isicas and National Institute of Science and Technology for Complex Systems, Rua Xavier Sigaud 150, Rio de Janeiro 22290-180, RJ, Brazil}
\affiliation{Santa Fe Institute, 1399 Hyde Park Road, Santa Fe, NM 87501, USA}

\begin{abstract}
We consider correlated random variables $X_1,\dots,X_n$ taking values in $\{0,1\}$ such that, for any permutation $\pi$ of $\{1,\dots,n\}$, the random vectors $(X_1,\dots,X_n)$ and $(X_{\pi(1)},\dots,X_{\pi(n)})$ have the same distribution. This distribution, which was introduced by Rodr\'iguez \textit{et al.} (2008) and then generalized by Hanel \textit{et al.} (2009), is scale-invariant and depends on a real parameter $\nu>0$ ($\nu\to\infty$ implies independence). Putting $S_n=X_1+\cdots+X_n$, the distribution of $S_n-n/2$ approaches a $Q$-Gaussian distribution with compact support ($Q=1-1/(\nu-1)<1$) as $n$ increases, after appropriate scaling. In the present article, we show that the distribution of $S_n/n$ converges, as $n\to\infty$, to a beta distribution with both parameters equal to~$\nu$. In particular, the law of large numbers does {\it not} hold since, if $0\le x<1/2$, then $\pr(S_n/n\le x)$, which is the probability of the event $\{S_n/n\le x\}$ (large deviation), does not converges to zero as $n\to\infty$. For $x=0$ and every real $\nu>0$, we show that $\pr(S_n=0)$ decays to zero like a power law of the form $1/n^\nu$ with a subdominant term of the form $1/n^{\nu+1}$. If $0<x\le 1$ and $\nu>0$ is an integer, we show that we can analytically find upper and lower bounds for the difference between $\pr(S_n/n\le x)$ and its ($n\to\infty$) limit. We also show that these bounds vanish like a power law of the form $1/n$ with a subdominant term of the form $1/n^2$.
\end{abstract}


\maketitle

\section{Introduction}
Nonextensive statistical mechanics~\cite{Tsallis1988,Tsallis2009} is a generalization of the celebrated Bolt\-zmann-Gibbs theory that is based on a nonadditive entropy, usually noted $S_q$, which depends on a real parameter $q$ ($S_1$ is the Boltzmann-Gibbs entropy). $S_q$ entropy is concave (convex) if $q>0$ ($q<0$). Extremization of this entropy with appropriate constraints~\cite{CuradoTsallis1991,TsallisMendesPlastino1998} yields $q$-exponentials, $f_q(y)\propto e_q^{-\lambda y}:=[1-(1-q)\lambda y]^{1/(1-q)}$ ($\lambda>0$), or $q$-Gaussians, $G_q(y)\propto e_q^{-\beta y^2}$ ($\beta>0$).~\cite{mathematica1,mathematica2} Both distributions appear in a large number of natural, artificial and social systems. For instance, in long-ranged-interacting many-body classical Hamiltonian systems,~\cite{PluchinoRapisardaTsallis2007,CirtoAssisTsallis2014,ChristodoulidiTsallisBountis2014} cold atoms in dissipative optical lattices,~\cite{DouglasBergaminiRenzoni2006} dusty plasmas,~\cite{LiuGoree2008} in the study of the over-damped motion of interacting particles,~\cite{AndradeSilvaMoreiraNobreCurado2010,RibeiroNobreCurado2012a,RibeiroNobreCurado2012b} in high energy physics~\cite{WongWilk2012,WongWilk2013,CirtoTsallisWongWilk2014} and in biology.~\cite{UpadhyayaRieuGlazierSawada2001}

A recent article~\cite{RuizTsallis2012} has illustrated that a generalized large deviation theory compatible with nonextensive statistical mechanics may exist. The authors of that article considered a probabilistic model for the toss of $n$ correlated coins in which the probability of obtaining $k$ heads is approximately given by a $Q$-Gaussian with $Q>1$ after appropriate scaling and centering (the approximation becomes better as larger values of $n$ are considered). Under those conditions, they have shown numerically that the probability of obtaining a number of heads not greater than $nx$ ($0\le x\le 1/2$) approaches zero like~$e_q^{-nr_q(x)}$ as $n$ increases, where $q$ depends on $Q$ and $r_q(x)$ is a non-negative function.

In the present article we consider a probabilistic model analogous to the one considered in Ref. \onlinecite{RuizTsallis2012}. The only difference is that here we consider a different distribution, namely the one that has been introduced by Rodr\'iguez \textit{et al.}~\cite{RodriguezSchwammleTsallis2008} as a generalization of the Leibnitz triangle (further generalized later on by Hanel \textit{et al.}~\cite{HanelThurnerTsallis2009}). This distribution yields also $Q$-Gaussians as limiting distributions, however with $Q<1$ (which corresponds to a compact support). As in Ref. \onlinecite{RuizTsallis2012}, we are interested in finding asymptotic expressions for the probability of large deviations, even when in our case the law of large numbers does not hold.

The organization of the article is as follows: In section~\ref{definitions} we define our model and the notation we will use. A physical interpretation of the model is also given in this section. In section~\ref{sec.llnfail} we show that the law of large numbers does not hold. In section~\ref{asympt} we obtain upper and lower bounds for the difference between the probability of large deviations and its limit. Finally, we conclude in section~\ref{conc}.

\section{Model}
\label{definitions}
In this article, $\pr(E)$ denote the probability of an event $E$. We will work with random variables $X_1,\dots,X_n$ taking values in $\{0,1\}$ such that for any permutation $\pi$ of $\{1,\dots,n\}$, the random vectors $(X_1,\dots,X_n)$ and $(X_{\pi(1)},\dots,X_{\pi(n)})$ have the same distribution.
In addition to that, the distribution of $S_n:=X_1+\cdots+X_n$ is given by~\cite{RodriguezSchwammleTsallis2008,HanelThurnerTsallis2009}
\beq
\label{probSn}
p_{\nu,n}(k):=\pr(S_n=k)=\binom{n}{k}\frac{B(\nu+k,\nu+n-k)}{B(\nu,\nu)}
\eeq
for $k=0,1,\dots,n$, where $\nu>0$ is a real parameter and
\beq
\label{beta}
B(a,b):=\int_0^1y^{a-1}(1-y)^{b-1}\,dy
\eeq
is the beta function, defined for all $a,b>0$. 

We can immediately verify the following properties:
\begin{enumerate}
\item $\sum_{k=0}^np_{\nu,n}(k)=1$.
\item The distribution given in~(\ref{probSn}) is scale-invariant since
\begin{multline}
\binom{n}{k}^{-1}p_{\nu,n}(k)=\binom{n+1}{k}^{-1}p_{\nu,{n+1}}(k)\\
+\binom{n+1}{k+1}^{-1}p_{\nu,{n+1}}(k+1)\,.
\end{multline}
\item The random variables $X_1,\dots,X_n$ are correlated but identically distributed since $\pr(X_i=1)=\pr(X_i=0)=1/2$ (marginal probabilities) for all $i=1,\dots,n$.
\item The expected value of $S_n$ is $n/2$.
\item $p_{\nu,n}(k)\to 1/2^n$ as $\nu\to\infty$, i.e. $X_1,\dots,X_n$ are independent and identically distributed random variables if $\nu\to\infty$.
\end{enumerate}

In order to give a physical interpretation to our model, we can think of a set of $n$ spins $1/2$. Here $M_i:=X_i-1/2$ ($i=1,\dots,n$) is the value of the projection on the $z$-axis of the $i$th spin and $\mathcal{M}_n:=M_1+\cdots+M_n=S_n-n/2$ is the total magnetic moment of the system. The spins are correlated since the distribution given in~(\ref{probSn}) is not binomial. Moreover, all the microscopic configurations yielding a chosen value for the total magnetic moment have the same probability.

We say that a random variable $X$ has a \textit{$q$-Gaussian distribution} with parameters $q\in(-\infty,3)$ and $\beta>0$ if it has density~\cite{UmarovTsallisSteinberg2008}
\beq
G_q(\beta,y):=\frac{\sqrt{\beta}}{N_q}\dsqr{1-(1-q)\beta y^2}_+^{1/(1-q)}\,,
\eeq
where $[y]_+:=y$ if $y>0$; otherwise $[y]_+:=0$, and
\beq
N_q:=\left\{
\begin{array}{ll}
\frac{2^{\frac{3-q}{1-q}}[\Gamma(\frac{2-q}{1-q})]^2}{\sqrt{1-q}\Gamma(\frac{2(2-q)}{1-q})}&\mbox{for }q<1\\
\sqrt{\pi}&\mbox{for }q=1\\
\frac{\sqrt{\pi}\Gamma(\frac{3-q}{2(q-1)})}{\sqrt{q-1}\Gamma(\frac{1}{q-1})}&\mbox{if }1<q<3\,.
\end{array}
\right.
\eeq
It can be noticed that, for $q<1$, the function $y\mapsto G_q(\beta,y)$ has compact support, namely the interval $[-\frac{1}{\sqrt{(1-q)\beta}},\frac{1}{\sqrt{(1-q)\beta}}]$; otherwise, if $1\le q<3$, the support of this function is the whole real line.

Rodr\'iguez \textit{et al}\cite{RodriguezSchwammleTsallis2008} have shown that, if $\nu>0$ is an integer, then
\beq
\label{approx}
np_{\nu,n}(k)\approx\frac{1}{B(\nu,\nu)}\dpar{\frac{k}{n}}^{\nu-1}\dpar{1-\frac{k}{n}}^{\nu-1}
\eeq
for large values of $n$. Later on, Hanel \textit{et al}\cite{HanelThurnerTsallis2009} have obtained this approximation for every real $\nu>0$. Turning back to our physical interpretation,~(\ref{approx}) implies that, for every real $\nu>1$, the distribution of the total magnetic moment of the system is approximated, after appropriate scaling, by a $Q$-Gaussian distribution with parameter $Q<1$ when the number of spins is very large. More precisely, we have that
\beq
\label{totalmom}
\pr(\mathcal{M}_n=m)\approx\frac{1}{n}G_Q\dpar{\frac{4}{1-Q}\,,\frac{m}{n}}
\eeq
for large values of $n$, where $-n/2\le m\le n/2$ and 
\beq
\label{Qnu}
Q:=1-\frac{1}{\nu -1}<1.
\eeq 

\section{The law of large numbers does not hold}
\label{sec.llnfail}
A sequence $Y_1,Y_2,\dots$ of random variables is said to be \textit{exchangeable} if, for every $n$ and any permutation $\pi$ of $\{1,\dots,n\}$, the distributions of $(Y_1,\dots,Y_n)$ and $(Y_{\pi(1)},\dots,Y_{\pi(n)})$ are the same.~\cite{Durrett} A classical result about exchangeable random variables is \textit{de Finetti's theorem}. A particular version of this theorem says that, if $Y_1,Y_2,\dots$ is a sequence of exchangeable random variables taking values in $\{0,1\}$, then there exists a distribution $F$ on $[0,1]$ such that~\cite{Durrett,FristedtGray}
\beq
\pr(Y_1+\cdots+Y_n=k)=\int_0^1\binom{n}{k}\theta^k(1-\theta)^{n-k}\,dF(\theta)\,.
\eeq
Moreover, 
\beq
\lim_{n\to\infty}\pr\dpar{\frac{Y_1+\cdots+Y_n}{n}\le y}=F(y)
\eeq
at every continuity point $y$ of $F$. Let us remark that if $F$ has a density $f$, then $F$ is continuous everywhere and
\beq
F(y)=\int_{-\infty}^yf(t)\,dt
\eeq
for every real $y$.

In our model, since the distribution of $(X_1,\dots,X_n)$ is scale invariant, we are working in fact with a sequence $X_1,X_2,\dots$ of exchangeable random variables. Then, de Finetti's theorem can be applied. Using~(\ref{beta}),~(\ref{probSn}) can be rewritten as
\beq
p_{\nu,n}(k)=\int_0^1\binom{n}{k}\theta^k(1-\theta)^{n-k}\dsqr{\frac{\theta^{\nu-1}(1-\theta)^{\nu-1}}{B(\nu,\nu)}}\,d\theta\,.
\eeq
The expression between brackets on the right hand side is the density of a beta distribution with both parameters equal to $\nu$. Then, by de Finetti's theorem,
\beq
\label{llnfail}
\lim_{n\to\infty}\pr\dpar{\frac{S_n}{n}\le x}=\int_0^x\frac{1}{B(\nu,\nu)}\theta^{\nu-1}(1-\theta)^{\nu-1}\,d\theta
\eeq
for every $x\in[0,1]$ (see the \hyperref[llnfail.proof]{Appendix} for an elementary proof of this fact, which does not use de Finetti's theorem, in the special case of integer $\nu>0$).

Fixed $x\in[0,1]$, let us define the function
\beq
F_{\nu,x}(n):=\pr\dpar{\frac{S_n}{n}\le x}=\sum_{k=0}^{\floor{nx}}p_{\nu,n}(k)\,,
\label{distSn/n}
\eeq
where $\floor{y}$ is the greatest integer not exceeding the real number~$y$, i.e. $\floor{y}\le y<\floor{y}+1$. If $0\le x<1/2$, the event $\{S_n/n\le x\}$ is called a large deviation since it consists of values of $S_n$ which are at a distance proportional to $n$ from its mean. If $0<x<1/2$, then, from~(\ref{llnfail}), it follows that the probability of large deviations $F_{\nu,x}(n)$ converges, as $n\to\infty$, to a limit different from zero. This implies that the law of large numbers does not hold.

Remembering the physical interpretation that we gave to our model in section~\ref{definitions},~(\ref{llnfail}) says that there is positive probability of having the arithmetic mean of magnetic moments different from zero. For instance, if $\nu>1$,~(\ref{llnfail}) implies that
\beq
\label{llnfail.mag}
\lim_{n\to\infty}\pr\dpar{\dabs{\frac{\mathcal{M}_n}{n}}>\mu}=1-\int_{-\mu}^\mu G_Q\dpar{\frac{4}{1-Q}\,,y}\,dy\,,
\eeq
where the parameter $Q$ was defined in~(\ref{Qnu}). The quantity in~(\ref{llnfail.mag}) is positive if $0\le \mu<1/2$. Nevertheless, the magnetization per particle is $0$, since the expected value of $\mathcal{M}_n/n$ converges to $0$ as $n\to\infty$. Therefore, replacing a random variable by its mean, which is something that is usually done in Boltzmann-Gibbs statistical mechanics when connecting say the canonical and microcanonical ensembles, is not necessarily admissible.

\section{Asymptotics}
\label{asympt}
Let us define the quantity
\beq
\label{distbeta}
\overline{F}_{\nu,x}:=\int_0^x\frac{1}{B(\nu,\nu)}y^{\nu-1}(1-y)^{\nu-1}\,dy
\eeq
for every $x\in[0,1]$. From~(\ref{llnfail}) we have that $F_{\nu,x}(n)\to \overline{F}_{\nu,x}$ as~$n\to\infty$. In this section, we are interested in finding how rapidly the function $\Delta_{\nu,x}(n):=F_{\nu,x}(n)-\overline{F}_{\nu,x}$ approaches zero as $n$ increases. We will assume in subsections~\ref{asympt.nu1},~\ref{asympt.nu2} and~\ref{asympt.nuint} (and also in the \hyperref[llnfail.proof]{Appendix}) that $n$ is a non-negative real number. The reason for this is that calculations of upper and lower bounds of $\Delta_{\nu,x}(n)$ can be found more easily when $n$ is not restricted to take positive integer values only. Moreover, if $\nu>0$ is an integer, it follows from~(\ref{distSn/n}) and~(\ref{probSn}) that
\begin{multline}
\label{delta.nuint}
\Delta_{\nu,x}(n)=\\
\frac{(2\nu-1)!\Gamma(n+1)}{[(\nu-1)!]^2\Gamma(2\nu+n)}\sum_{k=0}^{\floor{nx}}\prod_{j=1}^{\nu-1}(k+j)(n-k+j)-\overline{F}_{\nu,x}
\end{multline}
for every $x\in[0,1]$, and we notice that this expression is well-defined even for $n$ a non-negative real number.

\subsection{The case \texorpdfstring{$x=0$}{x=0}}
If $x=0$, then, using~(\ref{distSn/n}) and~(\ref{distbeta}), we have that $\Delta_{\nu,0}(n)=F_{\nu,0}(n)=p_{\nu,n}(0)$ for every real $\nu>0$. Then, using the relation
\beq
B(a,b)=\frac{\Gamma(a)\Gamma(b)}{\Gamma(a+b)}\,,
\eeq
which holds for all $a,b>0$, it follows immediately from~(\ref{probSn}) that
\beq
\Delta_{\nu,0}(n)=\frac{\Gamma(2\nu)\Gamma(\nu+n)}{\Gamma(\nu)\Gamma(2\nu+n)}
\eeq
for all real $\nu>0$ and integer $n\ge 1$.

We have that
\beq
\label{asympt.x0}
\Delta_{\nu,0}(n)\sim\frac{\Gamma(2\nu)}{\Gamma(\nu)}\frac{1}{n^\nu}\,,
\eeq
where the symbol $\sim$ (\textit{asymptotic equivalence}) means that the ratio of both sides tends to $1$ as $n\to\infty$. Indeed, by the Stirling formula (see theorem 8.22 in Ref.~\onlinecite{BabyRudin}),
\beq
\begin{split}
\frac{n^\nu\Gamma(\nu+n)}{\Gamma(2\nu+n)}&\sim\frac{\sqrt{2\pi}(\nu+n-1)^{\nu+n-1/2}e^{-\nu-n+1}n^\nu}{\sqrt{2\pi}(2\nu+n-1)^{2\nu+n-1/2}e^{-2\nu-n+1}}\\
&=\dpar{1-\frac{\nu}{2\nu+n-1}}^{2\nu+n-1}\dpar{\frac{ne}{2\nu+n-1}}^\nu\\
&\quad\times\dpar{1-\frac{\nu}{2\nu+n-1}}^{-\nu+1/2}\,,
\end{split}
\eeq
where the right hand side tends to $1$ as $n\to\infty$ since the first parenthesis tends to $e^{-\nu}$, the second one to $e^\nu$ and the last one to $1$ as $n\to\infty$.

Fixed a real number $q$, the \textit{$q$-exponential function} is defined by
\beq
e_q^x:=\left\{
\begin{array}{ll}
[1+(1-q)x]^{1/(1-q)}&\mbox{for }q\ne 1\\
e^x&\mbox{for }q=1
\end{array}
\right.
\eeq
for every $x$ such that $(q-1)x<1$. Using this function,~(\ref{asympt.x0}) can be rewritten~as
\beq
\Delta_{\nu,0}(n)\sim \frac{(q-1)^{1/(q-1)}\Gamma(\frac{2}{q-1})}{\Gamma(\frac{1}{q-1})}e_q^{-n}\,,
\eeq
where 
\beq
q =1+ \frac{1}{\nu} \,  \in(1,2) \,.
\eeq 
Then, from~(\ref{Qnu}), we obtain the following relation:
\beq
\label{q.Q.rel}
\frac{1}{q-1}=1+\frac{1}{1-Q}\,.
\eeq
This relation is similar to the one found heuristically in Ref.~\onlinecite{RuizTsallis2012}.

\begin{figure}[t]
\centering
\includegraphics[width=0.45\textwidth,keepaspectratio]{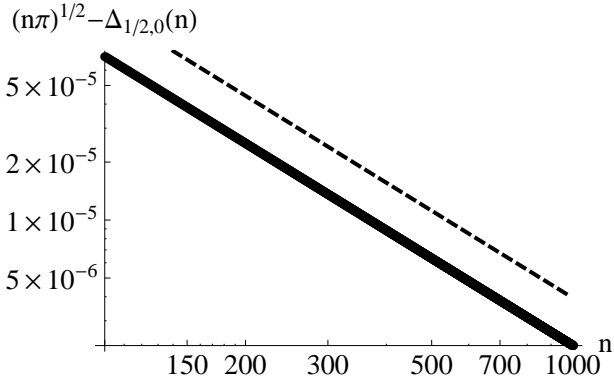}
\caption{Graph of the function $(n\pi)^{1/2}-\Delta_{1/2,0}(n)$ (solid line). The dashed line represents a power law of the form $n^{-3/2}$. The parallelism between the lines agrees with~(\ref{delta0.taylor}), which says that $\Delta_{1/2,0}(n)\approx \frac{1}{\sqrt{\pi n}}(1-\frac{1}{8n})$ for large~$n$.}
\label{deltah}
\end{figure}

For any integer $\nu>0$, it can be proved that
\beq
\label{subdom}
\lim_{n\to\infty}n^{\nu+1}\dsqr{\Delta_{\nu,0}(n)-\frac{\Gamma(2\nu)}{\Gamma(\nu)}\frac{1}{n^{\nu}}}=\frac{\nu(1-3\nu)}{2}\frac{\Gamma(2\nu)}{\Gamma(\nu)}\,.
\eeq
The proof follows straightforwardly by exploiting the fact that, for every integer $k\ge 2$,
\beq
\prod_{i=1}^k(n+i)=n^k+\frac{k(k+1)}{2}n^{k-1}+p_{k-2}(n)\,,
\eeq
where $p_{k-2}(n)$ is a polynomial in $n$ of degree $k-2$. A numerical study suggests that~(\ref{subdom}) is probably true for every real $\nu>0$ (for instance, see figure~\ref{deltah} for $\nu=1/2$). Therefore, in general we have
\beq
\label{delta0.taylor}
\Delta_{\nu,0}(n)=\frac{\Gamma(2\nu)}{\Gamma(\nu)}\frac{1}{n^\nu}\dcur{1-\frac{\nu(3\nu-1)}{2n}[1+h_\nu(n)]}\,,
\eeq
where $h_\nu(n)\to 0$ as $n\to\infty$. This expression is not precisely compatible with a $q$-ex\-po\-nen\-tial in the sense that the coefficients of the dominant and sub-dominant terms do not coincide with the corresponding ones of any function of the form $n\mapsto A(x)e_q^{-\lambda(x)n}$. This can be easily seen from the following expression, which is valid for every $q>1$:
\beq
\label{qexp.taylor}
e_q^{-\lambda n}=\frac{1}{[(q-1)\lambda n]^{1/(q-1)}}\dsqr{1-\frac{1}{(q-1)^2\lambda n}+h_q(n)}\,,
\eeq
where $nh_q(n)\to 0$ as $n\to\infty$.


\subsection{The case \texorpdfstring{$\nu=1$}{nu=1}}
\label{asympt.nu1}
\begin{figure}[t]
\centering
\includegraphics[width=0.45\textwidth,keepaspectratio]{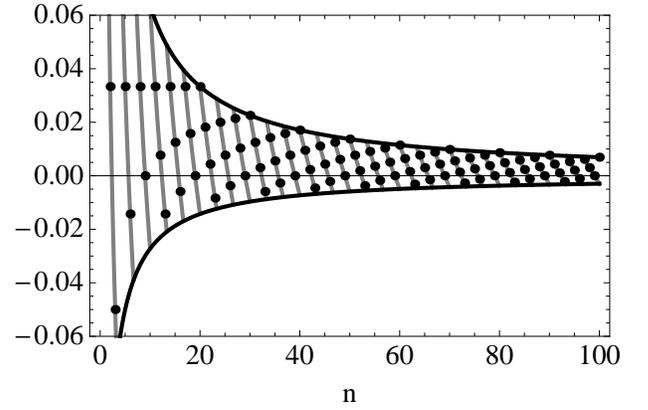}
\caption{Graphs of the functions $U_{1,3/10}(n)$ (upper solid curve) and $L_{1,3/10}(n)$ (lower solid curve). The graph of the function $\Delta_{1,3/10}(n)$ is the set of all segments within the two curves. The dots represent $\Delta_{1,3/10}(n)$ for integer values of~$n$.}
\label{delta1}
\end{figure}

Let us consider $\nu=1$ and $x\in[0,1]$. 
From~(\ref{distbeta}) we obtain $\overline{F}_{1,x}=x$. Then, from~(\ref{delta.nuint}) we have
\beq
\Delta_{1,x}(n)=\frac{\floor{nx}+1}{n+1}-x\,,
\eeq
where here $n$ is any non-negative real number. It can be verified immediately that
\beq
-\frac{x}{n+1}<\Delta_{1,x}(n)\le\frac{1-x}{n+1}\,.
\eeq
This means that the functions
\beq
\label{bounds1}
\begin{split}
U_{1,x}(n)&:=\frac{1-x}{n+1}=(1-x)e_2^{-n}\,,\\
L_{1,x}(n)&:=-\frac{x}{n+1}=-xe_2^{-n}
\end{split}
\eeq
are such that $L_{1,x}(n)<\Delta_{1,x}(n)\le U_{1,x}(n)$. In other words, $U_{1,x}(n)$ and $L_{1,x}(n)$ are respectively upper and lower bounds of $\Delta_{1,x}(n)$ (see figure~\ref{delta1}). Moreover, it can be seen immediately from~(\ref{bounds1}) that $L_{1,x}(n)=-U_{1,1-x}(n)$.

\subsection{The case \texorpdfstring{$\nu=2$}{nu=2}}
\label{asympt.nu2}
Let us consider $\nu=2$ and $x\in[0,1]$. From~(\ref{distbeta}) we have $\overline{F}_{2,x}=3x^2-2x^3$. Then, it can be obtained from~(\ref{delta.nuint}) that
\begin{multline}
\Delta_{2,n}(x)=\frac{3(\floor{nx}+1)(\floor{nx}+2)}{(n+2)(n+3)}\\
-\frac{2\floor{nx}(\floor{nx}+1)(\floor{nx}+2)}{(n+1)(n+2)(n+3)}-3x^2+2x^3\,,
\label{Delta2}
\end{multline}
where here $n$ is any non-negative real number. 

\begin{figure}[t]
\centering
\includegraphics[width=0.45\textwidth,keepaspectratio]{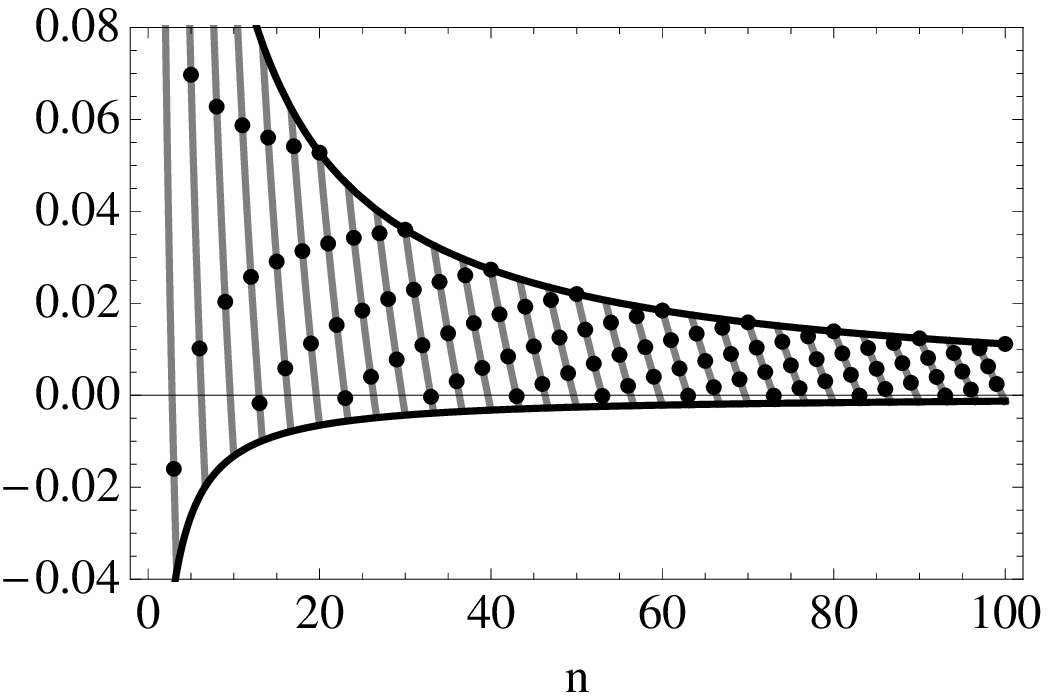}
\caption{Graph of the function $U_{2,3/10}(n)$ (upper solid curve) and $L_{2,3/10}(n)$ (lower solid curve). The graph of the function $\Delta_{2,3/10}(n)$ is the set of all segments within the two curves. The dots represent $\Delta_{2,3/10}(n)$ for integer values of~$n$.}
\label{delta2}
\end{figure}

If $x\ne 0$, the graph of the function $\Delta_{2,x}(n)$ (for instance, see figure~\ref{delta2} for $x=3/10$) shows that this function is monotonically decreasing in each interval $I_l:=[\frac{l}{x},\frac{l+1}{x})$, where $l\ge 0$ is an integer. Thus, for every integer $l\ge 0$, 
\beq
\begin{split}
\sup_{n\in I_l}\Delta_{2,x}(n)&=\Delta_{2,x}(l/x)\,,\\
\inf_{n\in I_l}\Delta_{2,x}(n)&=\lim_{n\to\frac{l+1}{x}-}\Delta_{2,x}(n)\,.
\end{split}
\eeq
Then, the functions
\begin{multline}
\label{U2}
U_{2,x}(n):=\frac{3(nx+1)(nx+2)}{(n+2)(n+3)}\\
-\frac{2nx(nx+1)(nx+2)}{(n+1)(n+2)(n+3)}-3x^2+2x^3
\end{multline}
and
\begin{multline}
\label{L2}
L_{2,x}(n):=\frac{3nx(nx+1)}{(n+2)(n+3)}\\
-\frac{2nx(nx-1)(nx+1)}{(n+1)(n+2)(n+3)}-3x^2+2x^3\,,
\end{multline}
are such that
\beq
\begin{split}
U_{2,x}(l/x)&=\sup_{n\in I_l}\Delta_{2,x}(n)\,,\\
L_{2,x}\dpar{\frac{l+1}{x}}&=\inf_{n\in I_l}\Delta_{2,x}(n)
\end{split}
\eeq
for every integer $l>0$. Moreover, figure~\ref{delta2} illustrates that $U_{2,x}(n)$ and $L_{2,x}(n)$ are respectively upper and lower bounds of $\Delta_{2,x}(n)$, i.e. $L_{2,x}(n)\le \Delta_{2,n}(n)\le U_{2,x}(n)$. As in the case of~$\nu=1$, it can be verified that $L_{2,x}(n)=-U_{2,1-x}(n)$.

We have seen that both bounds of the function $\Delta_{2,x}(n)$ are not proportional to $q$-ex\-po\-nen\-tials. Moreover, by Taylor's theorem, we have
\begin{multline}
U_{2,x}(n)=\frac{9x-21x^2+12x^3}{n}\\
+\frac{6-49x+93x^2-50x^3}{n^2}+h_U(n)
\end{multline}
and
\begin{multline}
L_{2,x}(n)=\frac{3x-15x^2+12x^3}{n}\\
-\frac{13x-57x^2+50x^3}{n^2}+h_L(n)\,,
\end{multline}
where $n^2h_U(n)\to 0$ and $n^2h_L(n)\to 0$ as $n\to\infty$. Both of these expressions are not precisely compatible with a $q$-exponential with $q=2$, in the sense that the coefficients of the dominant and sub-dominant terms do not coincide with the corresponding ones of any function of the form $n\mapsto A(x)e_2^{-\lambda(x)n}$ (see~(\ref{qexp.taylor})).

\subsection{The case of integer \texorpdfstring{$\nu>0$}{nu>0}}
\label{asympt.nuint}
In subsections~\ref{asympt.nu1} and~\ref{asympt.nu2} we have analytically found upper and lower bounds for the functions $\Delta_{1,x}(n)$ and $\Delta_{2,x}(n)$. We have seen that, in the cases $\nu=1$ and $\nu=2$, these upper and lower bounds approach zero like a power law of the form $1/n$ with a subdominant term of the form $1/n^2$. We claim that the procedure described in subsection~\ref{asympt.nu2} can be followed to obtain analytically upper and lower bounds for the function $\Delta_{\nu,x}(n)$ when $\nu$ is any positive integer.

The justification of our claim has two parts. The first part is that we can always find a closed expression for $\Delta_{\nu,x}(n)$ using~(\ref{delta.nuint}) when $\nu>0$ is an integer. This is because~(\ref{delta.nuint}) just involves sums of powers of the first $\floor{nx}$ positive integers, which have closed expressions. The second part of our justification is the hypothesis that the function $\Delta_{\nu,x}(n)$ is monotonically decreasing in each interval $[\frac{l}{x},\frac{l+1}{x})$, where $l\ge 0$ is an integer. This hypothesis has been verified to be correct for $\nu=1$ (analytically) and $\nu=2$ (numerically). Figures~\ref{delta3} and~\ref{delta10} illustrate that this hypothesis is also correct for $\nu=3$ and $\nu=10$ when $x=3/10$. We have verified that the same holds for other typical values of $x\in[0,1]$.

Now we give an illustration of the application of the procedure described in subsection~\ref{asympt.nu2} to the case $\nu=3$. Fixed $x\in[0,1]$, from~(\ref{delta.nuint}), it can be obtained that
\begin{widetext}
\begin{multline}
\label{delta3.expr}
\Delta_{3,x}(n)=\frac{(\floor{nx}+1)(\floor{nx}+2)(\floor{nx}+3)}{(n+1)(n+2)(n+3)(n+4)(n+5)}[10(n^2+3n+2)-3(5n+7)\floor{nx}+6\floor{nx}^2]\\
-6x^5+15x^4-10x^3\,,
\end{multline}

\begin{figure}[t]
\centering
\includegraphics[width=0.45\textwidth,keepaspectratio]{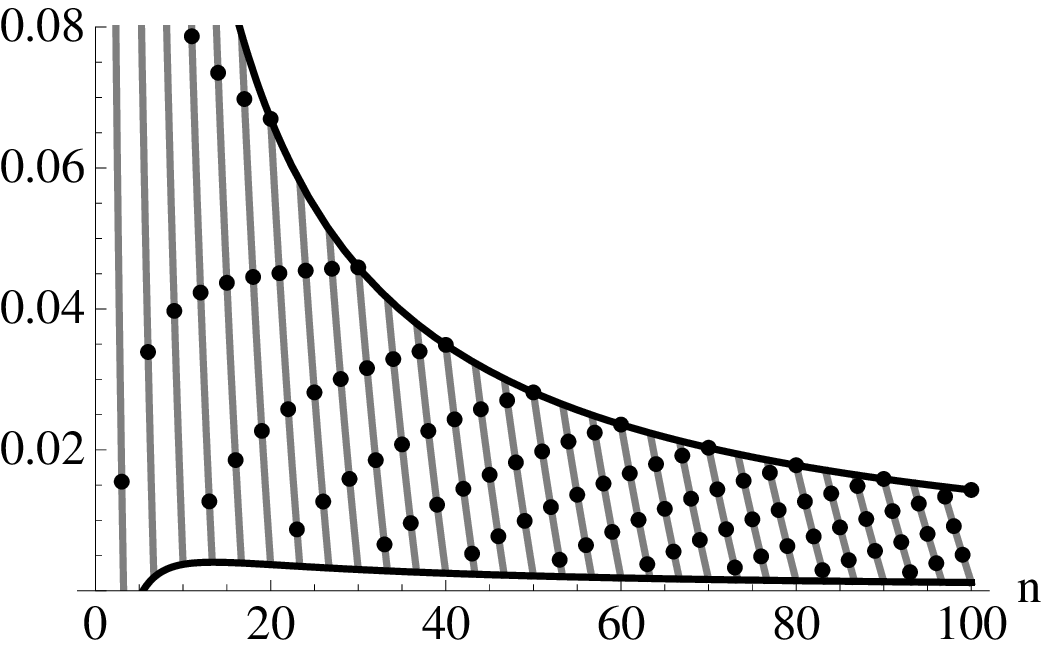}\quad\includegraphics[width=0.45\textwidth,keepaspectratio]{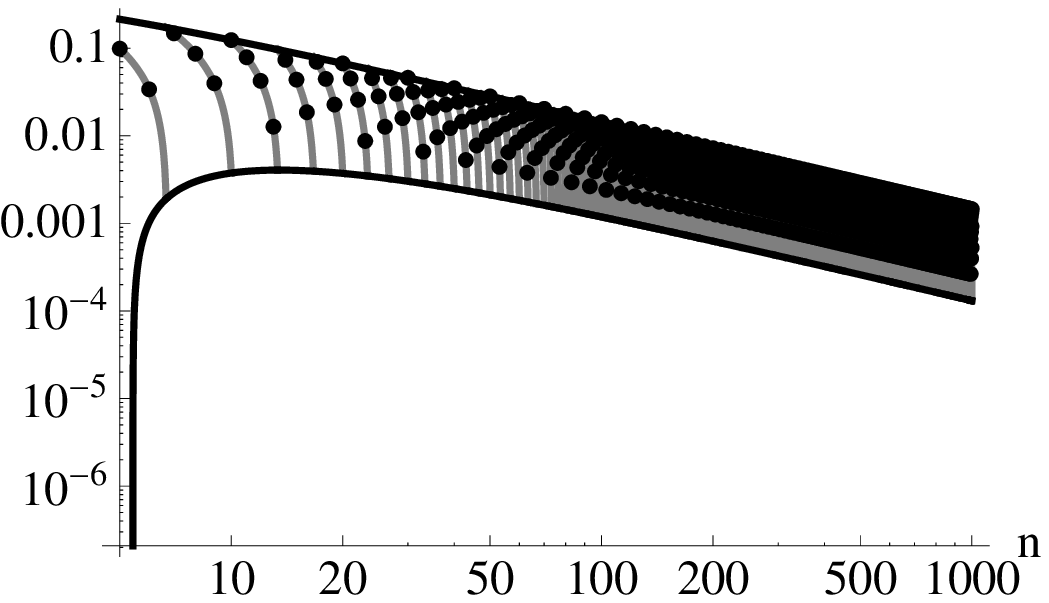}
\caption{(Left) The graph of the function $\Delta_{3,3/10}(n)$ is the set of all gray segments between the two solid curves, which are the graphs of its upper and lower bounds. The dots represent $\Delta_{3,3/10}(n)$ for integer values of~$n$. (Right) The same data represented in log-log scale in order to show the asymptotic power law behavior of the upper and lower bounds of $\Delta_{3,3/10}(n)$.}
\label{delta3}
\end{figure}

\begin{figure}[t]
\centering
\includegraphics[width=0.45\textwidth,keepaspectratio]{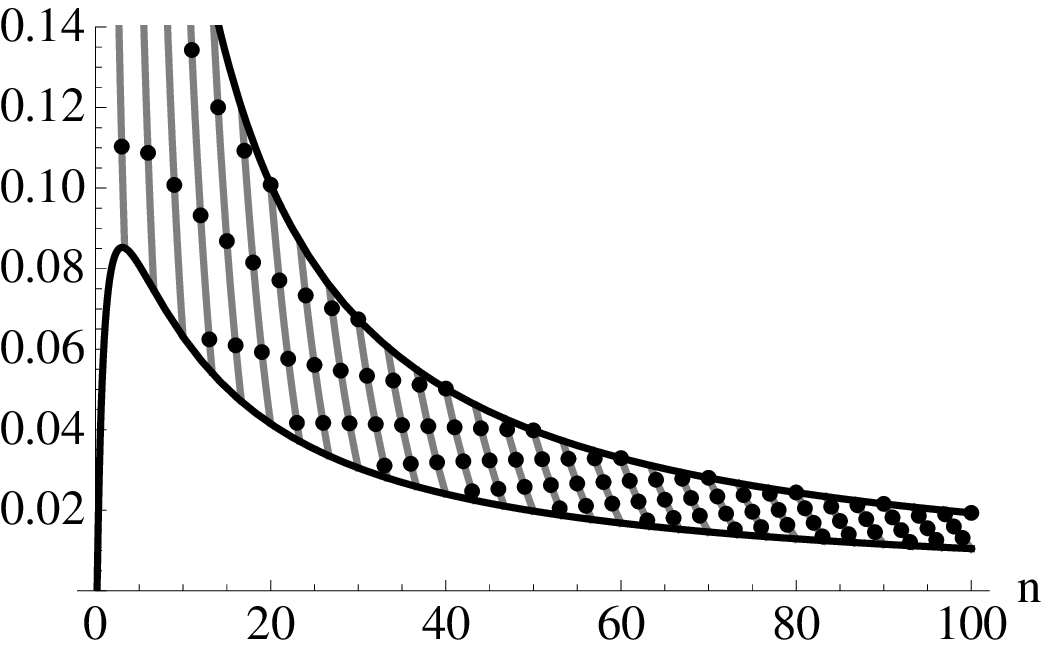}\quad\includegraphics[width=0.45\textwidth,keepaspectratio]{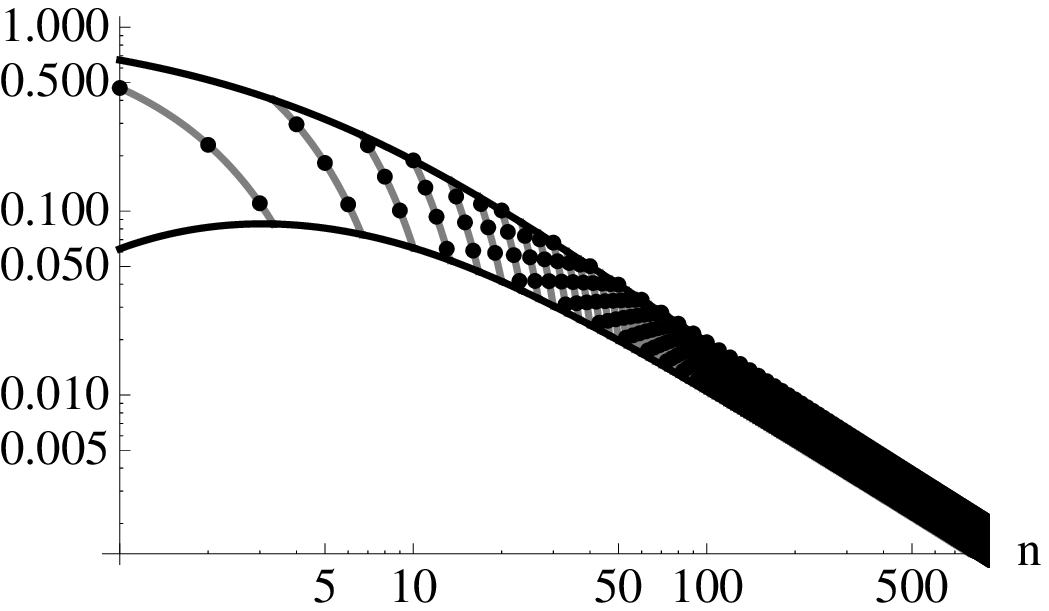}
\caption{(Left) The graph of the function $\Delta_{10,3/10}(n)$ is the set of all gray segments between the two solid curves, which are the graphs of its upper and lower bounds. The dots represent $\Delta_{10,3/10}(n)$ for integer values of~$n$. (Right) The same data represented in log-log scale in order to show the asymptotic power law behavior of the upper and lower bounds of $\Delta_{10,3/10}(n)$.}
\label{delta10}
\end{figure}
\end{widetext}
where $n$ is a non-negative real number. Then, the analytical expression for the upper bound of the function $\Delta_{3,x}(n)$ is obtained by replacing $\floor{nx}$ by $nx$ in~(\ref{delta3.expr}). For its lower bound, we have to replace $\floor{nx}$ by $nx-1$ in~(\ref{delta3.expr}). The graphs of the function $\Delta_{3,x}(n)$ and its bounds are shown in figure~\ref{delta3} when $x=3/10$. From the analytical expressions of the upper and lower bounds of the function $\Delta_{3,x}(n)$, it can be verified that, once again, they approach zero like a power law of the form $1/n$ with a subdominant term of the form $1/n^2$.

We also applied the procedure described in subsection~\ref{asympt.nu2} to the case $\nu=10$. However, in this case, the expressions are too cumbersome to be shown here. Nevertheless, the same conclusions as for the case $\nu=3$ are obtained (see figure~\ref{delta10}).

\section{Final remarks and conclusions}
\label{conc}
We have worked with a sequence $X_1,X_2,\dots$ of exchangeable and correlated random variables taking values in $\{0,1\}$ such that the distribution of the partial sums $S_n=X_1+\cdots+X_n$ depends on a real parameter $\nu>0$ (see~(\ref{probSn})). We have seen that the law of large numbers does not hold since, if $0<x<1/2$, the probability of the large deviation $\{S_n/n\le x\}$ converges, as $n\to\infty$, to a limit different from zero. As in Ref. \onlinecite{RuizTsallis2012}, here we were interested in studying how rapidly the probability of large deviations converges to its limit. 

In section~\ref{asympt} we have defined, for every $x\in[0,1]$, the function $\Delta_{\nu,x}(n)$ as being the difference between $\pr(S_n/n\le x)$ and its ($n\to\infty$) limit. We have seen that for $x=0$, the function $\Delta_{\nu,0}(n)$ decays to zero like a power law of the form $1/n^{\nu}$ with a subdominant term of the form $1/n^{\nu+1}$. If $0<x\le 1$, the results found in subsections~\ref{asympt.nu1},~\ref{asympt.nu2} and~\ref{asympt.nuint} allow us to conclude that, for each integer $\nu>0$, it may be possible to find analytical expressions for upper and lower bounds of $\Delta_{\nu,x}(n)$ such that they approach zero like a power law of the form $1/n$ with a subdominant term of the form $1/n^2$.

It seems interesting to remark that the model we have considered in this article yields $Q$-Gaussians as limiting distributions (see~(\ref{totalmom})).  However, with the exception of the case $\nu=1$, no other value of $\nu>0$ yields precisely $q$-exponential bounds for the function $\Delta_{\nu,x}(n)$, in contrast to the results found in the recent discussion of another model~\cite{RuizTsallis2013}.

As a matter of curiosity, we could investigate whether the law of large numbers holds or not when we consider a different distribution, which also yields $Q$-Gaussian limiting distributions with compact support ($Q<1$), instead of the one given in~(\ref{probSn}). Particularly, we can consider that~\cite{RodriguezSchwammleTsallis2008}
\begin{multline}
\label{discretized}
p^*_{Q,n}(k):=\pr(S_n=k)=\\
\frac{\dpar{\frac{k+1}{n+2}}^{1/(1-Q)}\dsqr{1-\dpar{\frac{k+1}{n+2}}}^{1/(1-Q)}}{\sum_{j=0}^n\dpar{\frac{j+1}{n+2}}^{1/(1-Q)}\dsqr{1-\dpar{\frac{j+1}{n+2}}}^{1/(1-Q)}}\,,
\end{multline}
where $Q<1$. It can be verified immediately that, if $\frac{k+1}{n+2}\to y$ as $n\to\infty$, then
\beq
\lim_{n\to\infty}(n+2)p^*_{Q,n}(k)=\frac{y^{1/(1-Q)}(1-y)^{1/(1-Q)}}{B\dpar{\frac{2-Q}{1-Q},\frac{2-Q}{1-Q}}}\,,
\eeq
which, through centering, can be shown to yield a $Q$-Gaussian distribution.

\begin{figure}[t]
\centering
\includegraphics[width=0.45\textwidth,keepaspectratio]{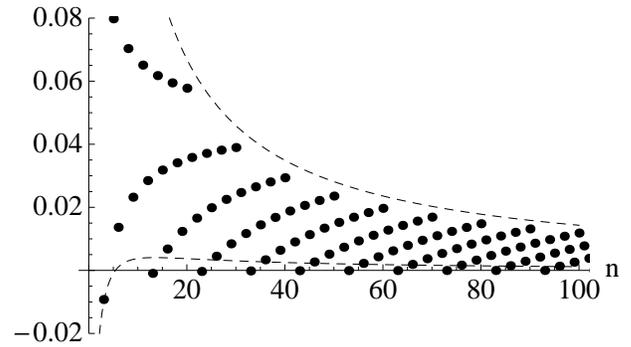}
\caption{Graph of the function $F^*_{1/2,3/10}(n)-\overline{F}_{3,3/10}$ (dots). Here we considered the distribution given in~(\ref{discretized}) with $Q=1/2$, instead of the one given in~(\ref{probSn}). The upper and lower dashed lines are respectively the upper and lower bounds of the function $\Delta_{3,x}(n)$.}
\label{fig.discretized}
\end{figure}

The distribution given in~(\ref{discretized}) is not scale-invariant for $Q\ne 0$ (for $Q=0$, $p^*_{0,n}(k)=p_{2,n}(k)$ according to~(\ref{probSn})). Hence, in this case we do not have a sequence $X_1,X_2,\dots$ of exchangeable random variables and, consequently, de Finetti's theorem does not hold. Nevertheless, we can verify numerically that the law of large numbers does {\it not} hold in this case either (see figure~\ref{fig.discretized}). Moreover, fixed $x\in[0,1]$, if $F^*_{Q,x}(n):=\pr(S_n/n\le x)$, then $F^*_{Q,x}(n)\to \overline{F}_{\nu,x}$ as $n\to\infty$, where $\overline{F}_{\nu,x}$ was defined in~(\ref{llnfail}) and $\nu$ can be obtained from~(\ref{Qnu}). However, the function $F^*_{Q,x}(n)-\overline{F}_{\nu,x}$ is not bounded by the same bounds we have found for the function $\Delta_{3,x}(n)$ (see section~\ref{asympt.nuint}).

Finally, we may conclude by saying that, for those systems where the correlations are not very strong (typically characterized by a Gaussian limiting distribution), both the law of large numbers and the large deviation theory hold (i.e., the probability of large deviations exponentially decays to zero when $n \to \infty$). In contrast, when strong correlations are present (e.g., characterized by $Q$-Gaussian limiting distributions), the law of large numbers can hold~\cite{RuizTsallis2012} or not (as in the present examples). In both strong correlated cases, the approach of the probability of large deviations  to its limit is a positive power law of $1/n$, in some particular cases not very different from a $q$-exponential behavior.

\section*{Acknowledgements}
We acknowledge L. J. L. Cirto, F. T. Lopes, A. R. Plastino, G. Ruiz and M.~E. Vares for fruitful discussions. Partial financial support from CNPq and Faperj (Brazilian agencies) also is acknowledged. One of us (CT) has also benefited from partial financial support from the John Templeton Foundation.

\appendix*
\section{Proof of~(\ref{llnfail}) in the case of integer \texorpdfstring{$\nu>0$}{nu>0}}
\label{llnfail.proof}
Let $\nu>0$ be an integer. Fixed $x\in[0,1]$, it can be obtained from~(\ref{distSn/n}) and~(\ref{probSn}) that
\begin{multline}
F_{\nu,x}(n)=\frac{(2\nu-1)!}{[(\nu-1)!]^2}\frac{1}{(n+1)\cdots(n+2\nu-1)}\\
\times\sum_{k=0}^{\floor{nx}}\prod_{j=1}^{\nu-1}(k+j)(n-k+j)\,.
\end{multline}
This expression is well-defined if $n$ is allowed to be any non-negative real number. We will prove that
\beq
\label{llnfail.cont}
\lim_{n\in[0,\infty),n\to\infty}F_{\nu,x}(n)=\int_0^x\frac{1}{B(\nu,\nu)}y^{\nu-1}(1-y)^{\nu-1}\,dy\,.
\eeq
This is a slightly stronger version of~(\ref{llnfail}) and is the correct version to be used in subsections~\ref{asympt.nu1},~\ref{asympt.nu2} and~\ref{asympt.nuint}, where we consider $n$ to assume non-negative real values.

If $x=0$,~(\ref{llnfail.cont}) is obviously true. The proof of~(\ref{llnfail.cont}) in the case of $x\in(0,1]$ relies on the two following simple facts:
\begin{enumerate}
\item For every integer $p\ge 1$, $1^p+\cdots+m^p$ is a polynomial in $m$ of degree $p+1$ with leading coefficient $1/(p+1)$. This can be easily proved using induction.
\item For every integer $p\ge 1$,
\beq
\lim_{y\to\infty}\frac{1}{y^{p+1}}\sum_{k=0}^{\floor{y}}k^p=\frac{1}{p+1}\,.
\eeq
This follows almost immediately from item 1.
\end{enumerate}

From item 1 we can conclude that
\beq
F_{\nu,x}(n)\sim\frac{(2\nu-1)!}{[(\nu-1)!]^2}\frac{1}{n^{2\nu-1}}\sum_{k=0}^{\floor{nx}}k^{\nu-1}(n-k)^{\nu-1}\,,
\eeq
where the symbol $\sim$ means that the ratio of both sides tends to $1$ as $n\to\infty$.
Then, by the binomial theorem,
\beq
F_{\nu,x}(n)\sim\frac{(2\nu-1)!}{[(\nu-1)!]^2}\sum_{j=0}^{\nu-1}\binom{\nu-1}{j}\frac{(-1)^j}{n^{\nu+j}}\sum_{k=0}^{\floor{nx}}k^{\nu+j-1}\,.
\eeq
Now, by item 2, the expression on the right hand side tends to
\beq
\frac{(2\nu-1)!}{[(\nu-1)!]^2}\sum_{j=0}^{\nu-1}\binom{\nu-1}{j}\frac{(-1)^jx^{\nu+j}}{\nu+j}
\eeq
as $n\to\infty$. Therefore,
\begin{multline}
\lim_{n\in[0,\infty),n\to\infty}F_{\nu,x}(n)=\\
\frac{(2\nu-1)!}{[(\nu-1)!]^2}\sum_{j=0}^{\nu-1}\binom{\nu-1}{j}\frac{(-1)^jx^{\nu+j}}{\nu+j}\,.
\end{multline}
On the other hand, as $\nu>0$ is an integer, it can be obtained that
\begin{multline}
\int_0^x\frac{1}{B(\nu,\nu)}y^{\nu-1}(1-y)^{\nu-1}\,dy=\\
\frac{(2\nu-1)!}{[(\nu-1)!]^2}\sum_{j=0}^{\nu-1}\binom{\nu-1}{j}\frac{(-1)^jx^{\nu+j}}{\nu+j}\,,
\end{multline}
where we have used the binomial theorem in the integrand on the left hand side. Comparing the last two expressions yields~(\ref{llnfail.cont}).

\bibliographystyle{aipnum4-1}
\bibliography{articles,books}

\begin{thebibliography}{26}%
\makeatletter
\providecommand \@ifxundefined [1]{%
 \@ifx{#1\undefined}
}%
\providecommand \@ifnum [1]{%
 \ifnum #1\expandafter \@firstoftwo
 \else \expandafter \@secondoftwo
 \fi
}%
\providecommand \@ifx [1]{%
 \ifx #1\expandafter \@firstoftwo
 \else \expandafter \@secondoftwo
 \fi
}%
\providecommand \natexlab [1]{#1}%
\providecommand \enquote  [1]{``#1''}%
\providecommand \bibnamefont  [1]{#1}%
\providecommand \bibfnamefont [1]{#1}%
\providecommand \citenamefont [1]{#1}%
\providecommand \href@noop [0]{\@secondoftwo}%
\providecommand \href [0]{\begingroup \@sanitize@url \@href}%
\providecommand \@href[1]{\@@startlink{#1}\@@href}%
\providecommand \@@href[1]{\endgroup#1\@@endlink}%
\providecommand \@sanitize@url [0]{\catcode `\\12\catcode `\$12\catcode
  `\&12\catcode `\#12\catcode `\^12\catcode `\_12\catcode `\%12\relax}%
\providecommand \@@startlink[1]{}%
\providecommand \@@endlink[0]{}%
\providecommand \url  [0]{\begingroup\@sanitize@url \@url }%
\providecommand \@url [1]{\endgroup\@href {#1}{\urlprefix }}%
\providecommand \urlprefix  [0]{URL }%
\providecommand \Eprint [0]{\href }%
\providecommand \doibase [0]{http://dx.doi.org/}%
\providecommand \selectlanguage [0]{\@gobble}%
\providecommand \bibinfo  [0]{\@secondoftwo}%
\providecommand \bibfield  [0]{\@secondoftwo}%
\providecommand \translation [1]{[#1]}%
\providecommand \BibitemOpen [0]{}%
\providecommand \bibitemStop [0]{}%
\providecommand \bibitemNoStop [0]{.\EOS\space}%
\providecommand \EOS [0]{\spacefactor3000\relax}%
\providecommand \BibitemShut  [1]{\csname bibitem#1\endcsname}%
\let\auto@bib@innerbib\@empty
\bibitem [{\citenamefont {Tsallis}(1988)}]{Tsallis1988}%
  \BibitemOpen
  \bibfield  {author} {\bibinfo {author} {\bibfnamefont {C.}~\bibnamefont
  {Tsallis}},\ }\href@noop {} {\bibfield  {journal} {\bibinfo  {journal} {J.
  Stat. Phys.}\ }\textbf {\bibinfo {volume} {52}},\ \bibinfo {pages} {479}
  (\bibinfo {year} {1988})}\BibitemShut {NoStop}%
\bibitem [{\citenamefont {Tsallis}(2009)}]{Tsallis2009}%
  \BibitemOpen
  \bibfield  {author} {\bibinfo {author} {\bibfnamefont {C.}~\bibnamefont
  {Tsallis}},\ }\href@noop {} {\emph {\bibinfo {title} {{I}ntroduction to
  {N}onextensive {S}tatistical {M}echanics: {A}pproaching a {C}omplex
  {W}orld}}}\ (\bibinfo  {publisher} {Springer},\ \bibinfo {year}
  {2009})\BibitemShut {NoStop}%
\bibitem [{\citenamefont {Curado}\ and\ \citenamefont
  {Tsallis}(1991)}]{CuradoTsallis1991}%
  \BibitemOpen
  \bibfield  {author} {\bibinfo {author} {\bibfnamefont {E.~M.~F.}\
  \bibnamefont {Curado}}\ and\ \bibinfo {author} {\bibfnamefont
  {C.}~\bibnamefont {Tsallis}},\ }\href@noop {} {\bibfield  {journal} {\bibinfo
   {journal} {J. Phys. A}\ }\textbf {\bibinfo {volume} {24}} (\bibinfo {year}
  {1991})}\BibitemShut {NoStop}%
\bibitem [{\citenamefont {Tsallis}, \citenamefont {Mendes},\ and\ \citenamefont
  {Plastino}(1998)}]{TsallisMendesPlastino1998}%
  \BibitemOpen
  \bibfield  {author} {\bibinfo {author} {\bibfnamefont {C.}~\bibnamefont
  {Tsallis}}, \bibinfo {author} {\bibfnamefont {R.~S.}\ \bibnamefont {Mendes}},
  \ and\ \bibinfo {author} {\bibfnamefont {A.~R.}\ \bibnamefont {Plastino}},\
  }\href@noop {} {\bibfield  {journal} {\bibinfo  {journal} {Physica A}\
  }\textbf {\bibinfo {volume} {261}},\ \bibinfo {pages} {534} (\bibinfo {year}
  {1998})}\BibitemShut {NoStop}%
\bibitem [{mat(2014{\natexlab{a}})}]{mathematica1}%
  \BibitemOpen
  \href@noop {} {\enquote {\bibinfo {title}
  {{T}sallis{Q}{G}aussian{D}istribution -- {W}olfram {L}anguage
  {D}ocumentation},}\ } (\bibinfo {year} {2014}{\natexlab{a}}),\ \bibinfo
  {note}
  {{\\\tiny\url{http://reference.wolfram.com/mathematica/ref/TsallisQGaussianDistribution.html}}}\BibitemShut
  {NoStop}%
\bibitem [{mat(2014{\natexlab{b}})}]{mathematica2}%
  \BibitemOpen
  \href@noop {} {\enquote {\bibinfo {title}
  {{T}sallis{Q}{E}xponential{D}istribution -- {W}olfram {L}anguage
  {D}ocumentation},}\ } (\bibinfo {year} {2014}{\natexlab{b}}),\ \bibinfo
  {note}
  {{\\\tiny\url{http://reference.wolfram.com/mathematica/ref/TsallisQExponentialDistribution.html}}}\BibitemShut
  {NoStop}%
\bibitem [{\citenamefont {Pluchino}, \citenamefont {Rapisarda},\ and\
  \citenamefont {Tsallis}(2007)}]{PluchinoRapisardaTsallis2007}%
  \BibitemOpen
  \bibfield  {author} {\bibinfo {author} {\bibfnamefont {A.}~\bibnamefont
  {Pluchino}}, \bibinfo {author} {\bibfnamefont {A.}~\bibnamefont {Rapisarda}},
  \ and\ \bibinfo {author} {\bibfnamefont {C.}~\bibnamefont {Tsallis}},\
  }\href@noop {} {\bibfield  {journal} {\bibinfo  {journal} {EPL}\ }\textbf
  {\bibinfo {volume} {80}},\ \bibinfo {pages} {26002} (\bibinfo {year}
  {2007})}\BibitemShut {NoStop}%
\bibitem [{\citenamefont {Cirto}, \citenamefont {Assis},\ and\ \citenamefont
  {Tsallis}(2014)}]{CirtoAssisTsallis2014}%
  \BibitemOpen
  \bibfield  {author} {\bibinfo {author} {\bibfnamefont {L.~J.~L.}\
  \bibnamefont {Cirto}}, \bibinfo {author} {\bibfnamefont {V.~R.~V.}\
  \bibnamefont {Assis}}, \ and\ \bibinfo {author} {\bibfnamefont
  {C.}~\bibnamefont {Tsallis}},\ }\href@noop {} {\bibfield  {journal} {\bibinfo
   {journal} {Physica A}\ }\textbf {\bibinfo {volume} {393}} (\bibinfo {year}
  {2014})}\BibitemShut {NoStop}%
\bibitem [{\citenamefont {Christodoulidi}, \citenamefont {Tsallis},\ and\
  \citenamefont {Bountis}(2014)}]{ChristodoulidiTsallisBountis2014}%
  \BibitemOpen
  \bibfield  {author} {\bibinfo {author} {\bibfnamefont {H.}~\bibnamefont
  {Christodoulidi}}, \bibinfo {author} {\bibfnamefont {C.}~\bibnamefont
  {Tsallis}}, \ and\ \bibinfo {author} {\bibfnamefont {T.}~\bibnamefont
  {Bountis}},\ }\href@noop {} {\enquote {\bibinfo {title}
  {{F}ermi-{P}asta-{U}lam model with long-range interactions: Dynamics and
  thermostatistics},}\ } (\bibinfo {year} {2014}),\ \Eprint
  {http://arxiv.org/abs/1405.3528} {arXiv:1405.3528} \BibitemShut {NoStop}%
\bibitem [{\citenamefont {Douglas}, \citenamefont {Bergamini},\ and\
  \citenamefont {Renzoni}(2006)}]{DouglasBergaminiRenzoni2006}%
  \BibitemOpen
  \bibfield  {author} {\bibinfo {author} {\bibfnamefont {P.}~\bibnamefont
  {Douglas}}, \bibinfo {author} {\bibfnamefont {S.}~\bibnamefont {Bergamini}},
  \ and\ \bibinfo {author} {\bibfnamefont {F.}~\bibnamefont {Renzoni}},\
  }\href@noop {} {\bibfield  {journal} {\bibinfo  {journal} {Phys. Rev. Lett.}\
  }\textbf {\bibinfo {volume} {96}},\ \bibinfo {pages} {110601} (\bibinfo
  {year} {2006})}\BibitemShut {NoStop}%
\bibitem [{\citenamefont {Liu}\ and\ \citenamefont
  {Goree}(2008)}]{LiuGoree2008}%
  \BibitemOpen
  \bibfield  {author} {\bibinfo {author} {\bibfnamefont {B.}~\bibnamefont
  {Liu}}\ and\ \bibinfo {author} {\bibfnamefont {J.}~\bibnamefont {Goree}},\
  }\href@noop {} {\bibfield  {journal} {\bibinfo  {journal} {Phys. Rev. Lett.}\
  }\textbf {\bibinfo {volume} {100}},\ \bibinfo {pages} {055003} (\bibinfo
  {year} {2008})}\BibitemShut {NoStop}%
\bibitem [{\citenamefont {{Andrade Jr.}}\ \emph {et~al.}(2010)\citenamefont
  {{Andrade Jr.}}, \citenamefont {{da Silva}}, \citenamefont {Moreira},
  \citenamefont {Nobre},\ and\ \citenamefont
  {Curado}}]{AndradeSilvaMoreiraNobreCurado2010}%
  \BibitemOpen
  \bibfield  {author} {\bibinfo {author} {\bibfnamefont {J.~S.}\ \bibnamefont
  {{Andrade Jr.}}}, \bibinfo {author} {\bibfnamefont {G.~F.~T.}\ \bibnamefont
  {{da Silva}}}, \bibinfo {author} {\bibfnamefont {A.~A.}\ \bibnamefont
  {Moreira}}, \bibinfo {author} {\bibfnamefont {F.~D.}\ \bibnamefont {Nobre}},
  \ and\ \bibinfo {author} {\bibfnamefont {E.~M.~F.}\ \bibnamefont {Curado}},\
  }\href@noop {} {\bibfield  {journal} {\bibinfo  {journal} {Phys. Rev. Lett.}\
  }\textbf {\bibinfo {volume} {105}},\ \bibinfo {pages} {260601} (\bibinfo
  {year} {2010})}\BibitemShut {NoStop}%
\bibitem [{\citenamefont {Ribeiro}, \citenamefont {Nobre},\ and\ \citenamefont
  {Curado}(2012{\natexlab{a}})}]{RibeiroNobreCurado2012a}%
  \BibitemOpen
  \bibfield  {author} {\bibinfo {author} {\bibfnamefont {M.~S.}\ \bibnamefont
  {Ribeiro}}, \bibinfo {author} {\bibfnamefont {F.~D.}\ \bibnamefont {Nobre}},
  \ and\ \bibinfo {author} {\bibfnamefont {E.~M.~F.}\ \bibnamefont {Curado}},\
  }\href@noop {} {\bibfield  {journal} {\bibinfo  {journal} {Eur. Phys. J. B}\
  }\textbf {\bibinfo {volume} {85}},\ \bibinfo {pages} {399} (\bibinfo {year}
  {2012}{\natexlab{a}})}\BibitemShut {NoStop}%
\bibitem [{\citenamefont {Ribeiro}, \citenamefont {Nobre},\ and\ \citenamefont
  {Curado}(2012{\natexlab{b}})}]{RibeiroNobreCurado2012b}%
  \BibitemOpen
  \bibfield  {author} {\bibinfo {author} {\bibfnamefont {M.~S.}\ \bibnamefont
  {Ribeiro}}, \bibinfo {author} {\bibfnamefont {F.~D.}\ \bibnamefont {Nobre}},
  \ and\ \bibinfo {author} {\bibfnamefont {E.~M.~F.}\ \bibnamefont {Curado}},\
  }\href@noop {} {\bibfield  {journal} {\bibinfo  {journal} {Phys. Rev. E}\
  }\textbf {\bibinfo {volume} {85}},\ \bibinfo {pages} {021146} (\bibinfo
  {year} {2012}{\natexlab{b}})}\BibitemShut {NoStop}%
\bibitem [{\citenamefont {Wong}\ and\ \citenamefont
  {Wilk}(2012)}]{WongWilk2012}%
  \BibitemOpen
  \bibfield  {author} {\bibinfo {author} {\bibfnamefont {C.~Y.}\ \bibnamefont
  {Wong}}\ and\ \bibinfo {author} {\bibfnamefont {G.}~\bibnamefont {Wilk}},\
  }\href@noop {} {\bibfield  {journal} {\bibinfo  {journal} {Acta Phys.
  Polonica B}\ }\textbf {\bibinfo {volume} {43}},\ \bibinfo {pages} {2047}
  (\bibinfo {year} {2012})}\BibitemShut {NoStop}%
\bibitem [{\citenamefont {Wong}\ and\ \citenamefont
  {Wilk}(2013)}]{WongWilk2013}%
  \BibitemOpen
  \bibfield  {author} {\bibinfo {author} {\bibfnamefont {C.~Y.}\ \bibnamefont
  {Wong}}\ and\ \bibinfo {author} {\bibfnamefont {G.}~\bibnamefont {Wilk}},\
  }\href@noop {} {\bibfield  {journal} {\bibinfo  {journal} {Acta Phys.
  Polonica B}\ }\textbf {\bibinfo {volume} {87}},\ \bibinfo {pages} {114007}
  (\bibinfo {year} {2013})}\BibitemShut {NoStop}%
\bibitem [{\citenamefont {Cirto}\ \emph {et~al.}(2014)\citenamefont {Cirto},
  \citenamefont {Tsallis}, \citenamefont {Wong},\ and\ \citenamefont
  {Wilk}}]{CirtoTsallisWongWilk2014}%
  \BibitemOpen
  \bibfield  {author} {\bibinfo {author} {\bibfnamefont {L.~J.~L.}\
  \bibnamefont {Cirto}}, \bibinfo {author} {\bibfnamefont {C.}~\bibnamefont
  {Tsallis}}, \bibinfo {author} {\bibfnamefont {C.~Y.}\ \bibnamefont {Wong}}, \
  and\ \bibinfo {author} {\bibfnamefont {G.}~\bibnamefont {Wilk}},\ }\href@noop
  {} {\enquote {\bibinfo {title} {The transverse-momenta distributions in
  high-energy $pp$ collisions -- {A} statistical-mechanical approach},}\ }
  (\bibinfo {year} {2014}),\ \Eprint {http://arxiv.org/abs/1409.3278}
  {arXiv:1409.3278} \BibitemShut {NoStop}%
\bibitem [{\citenamefont {Upadhyaya}\ \emph {et~al.}(2001)\citenamefont
  {Upadhyaya}, \citenamefont {Rieu}, \citenamefont {Glazier},\ and\
  \citenamefont {Sawada}}]{UpadhyayaRieuGlazierSawada2001}%
  \BibitemOpen
  \bibfield  {author} {\bibinfo {author} {\bibfnamefont {A.}~\bibnamefont
  {Upadhyaya}}, \bibinfo {author} {\bibfnamefont {J.~P.}\ \bibnamefont {Rieu}},
  \bibinfo {author} {\bibfnamefont {J.~A.}\ \bibnamefont {Glazier}}, \ and\
  \bibinfo {author} {\bibfnamefont {Y.}~\bibnamefont {Sawada}},\ }\href@noop {}
  {\bibfield  {journal} {\bibinfo  {journal} {Physica A}\ }\textbf {\bibinfo
  {volume} {293}},\ \bibinfo {pages} {549} (\bibinfo {year}
  {2001})}\BibitemShut {NoStop}%
\bibitem [{\citenamefont {Ruiz}\ and\ \citenamefont
  {Tsallis}(2012)}]{RuizTsallis2012}%
  \BibitemOpen
  \bibfield  {author} {\bibinfo {author} {\bibfnamefont {G.}~\bibnamefont
  {Ruiz}}\ and\ \bibinfo {author} {\bibfnamefont {C.}~\bibnamefont {Tsallis}},\
  }\href@noop {} {\bibfield  {journal} {\bibinfo  {journal} {Phys. Lett. A}\
  }\textbf {\bibinfo {volume} {376}},\ \bibinfo {pages} {2451} (\bibinfo {year}
  {2012})}\BibitemShut {NoStop}%
\bibitem [{\citenamefont {Rodr{\'i}guez}, \citenamefont {Schw{\"a}mmle},\ and\
  \citenamefont {Tsallis}(2008)}]{RodriguezSchwammleTsallis2008}%
  \BibitemOpen
  \bibfield  {author} {\bibinfo {author} {\bibfnamefont {A.}~\bibnamefont
  {Rodr{\'i}guez}}, \bibinfo {author} {\bibfnamefont {V.}~\bibnamefont
  {Schw{\"a}mmle}}, \ and\ \bibinfo {author} {\bibfnamefont {C.}~\bibnamefont
  {Tsallis}},\ }\href@noop {} {\bibfield  {journal} {\bibinfo  {journal} {J.
  Stat. Mech.}\ ,\ \bibinfo {pages} {P09006}} (\bibinfo {year}
  {2008})}\BibitemShut {NoStop}%
\bibitem [{\citenamefont {Hanel}, \citenamefont {Thurner},\ and\ \citenamefont
  {Tsallis}(2009)}]{HanelThurnerTsallis2009}%
  \BibitemOpen
  \bibfield  {author} {\bibinfo {author} {\bibfnamefont {R.}~\bibnamefont
  {Hanel}}, \bibinfo {author} {\bibfnamefont {S.}~\bibnamefont {Thurner}}, \
  and\ \bibinfo {author} {\bibfnamefont {C.}~\bibnamefont {Tsallis}},\
  }\href@noop {} {\bibfield  {journal} {\bibinfo  {journal} {Eur. Phys. J. B}\
  }\textbf {\bibinfo {volume} {72}},\ \bibinfo {pages} {263} (\bibinfo {year}
  {2009})}\BibitemShut {NoStop}%
\bibitem [{\citenamefont {Umarov}, \citenamefont {Tsallis},\ and\ \citenamefont
  {Steinberg}(2008)}]{UmarovTsallisSteinberg2008}%
  \BibitemOpen
  \bibfield  {author} {\bibinfo {author} {\bibfnamefont {S.}~\bibnamefont
  {Umarov}}, \bibinfo {author} {\bibfnamefont {C.}~\bibnamefont {Tsallis}}, \
  and\ \bibinfo {author} {\bibfnamefont {S.}~\bibnamefont {Steinberg}},\
  }\href@noop {} {\bibfield  {journal} {\bibinfo  {journal} {Milan J. Math.}\
  }\textbf {\bibinfo {volume} {76}},\ \bibinfo {pages} {307} (\bibinfo {year}
  {2008})}\BibitemShut {NoStop}%
\bibitem [{\citenamefont {Durrett}(2010)}]{Durrett}%
  \BibitemOpen
  \bibfield  {author} {\bibinfo {author} {\bibfnamefont {R.}~\bibnamefont
  {Durrett}},\ }\href@noop {} {\emph {\bibinfo {title} {Probability: {T}heory
  and {E}xamples}}},\ \bibinfo {edition} {4th}\ ed.\ (\bibinfo  {publisher}
  {Cambridge University Press},\ \bibinfo {year} {2010})\BibitemShut {NoStop}%
\bibitem [{\citenamefont {Fristedt}\ and\ \citenamefont
  {Gray}(1996)}]{FristedtGray}%
  \BibitemOpen
  \bibfield  {author} {\bibinfo {author} {\bibfnamefont {B.~E.}\ \bibnamefont
  {Fristedt}}\ and\ \bibinfo {author} {\bibfnamefont {L.~F.}\ \bibnamefont
  {Gray}},\ }\href@noop {} {\emph {\bibinfo {title} {A {M}odern {A}pproach to
  {P}robability {T}heory}}}\ (\bibinfo  {publisher} {{Birkh\"auser}},\ \bibinfo
  {year} {1996})\BibitemShut {NoStop}%
\bibitem [{\citenamefont {Rudin}(1976)}]{BabyRudin}%
  \BibitemOpen
  \bibfield  {author} {\bibinfo {author} {\bibfnamefont {W.}~\bibnamefont
  {Rudin}},\ }\href@noop {} {\emph {\bibinfo {title} {Principles of
  {M}athematical {A}nalysis}}},\ \bibinfo {edition} {3rd}\ ed.\ (\bibinfo
  {publisher} {McGraw-Hill},\ \bibinfo {year} {1976})\BibitemShut {NoStop}%
\bibitem [{\citenamefont {Ruiz}\ and\ \citenamefont
  {Tsallis}(2013)}]{RuizTsallis2013}%
  \BibitemOpen
  \bibfield  {author} {\bibinfo {author} {\bibfnamefont {G.}~\bibnamefont
  {Ruiz}}\ and\ \bibinfo {author} {\bibfnamefont {C.}~\bibnamefont {Tsallis}},\
  }\href@noop {} {\bibfield  {journal} {\bibinfo  {journal} {Phys. Lett. A}\
  }\textbf {\bibinfo {volume} {377}},\ \bibinfo {pages} {491} (\bibinfo {year}
  {2013})}\BibitemShut {NoStop}%
\end{thebibliography}%
\end{document}